\documentclass[twocolumn,A4]{article}
\usepackage[dvips]{graphics}
\usepackage{setspace}
\usepackage{color}
\definecolor{gold}{rgb}{0.85,0.66,0}
\definecolor{dblue}{rgb}{0,0,0.5}
\begin{document}
\onecolumn
\begin{center}
{\bf{\Large {\textcolor{gold}{Electronic transport in a mesoscopic ring}}}}\\
~\\
{\textcolor{dblue}{Santanu K. Maiti}}$^{1,2,*}$ \\
~\\
{\em $^1$Theoretical Condensed Matter Physics Division,
Saha Institute of Nuclear Physics, \\
1/AF, Bidhannagar, Kolkata-700 064, India \\
$^2$Department of Physics, Narasinha Dutt College,
129, Belilious Road, Howrah-711 101, India} \\
~\\
{\bf Abstract}
\end{center}
Electron transport properties of a non-interacting mesoscopic ring
sandwiched between two metallic electrodes are investigated by the use of 
Green's function formalism. We introduce a parametric approach based on 
the tight-binding model to study the transport properties. The electronic
transport characteristics are investigated in three aspects: (a) 
ring-electrode interface geometry, (b) coupling strength of the ring 
to the electrodes and (c) magnetic flux threaded by the ring.
\vskip 1cm
\begin{flushleft}
\end{flushleft}
\vskip 5.1in
\noindent
{\bf ~$^*$Corresponding Author}: Santanu K. Maiti

Electronic mail:  santanu.maiti@saha.ac.in
\newpage
\twocolumn

\section{Introduction}

Advanced progress of nanoscience and nanotechnology has allowed us 
to study the electron transport through mesoscopic rings in a very 
controllable way. Several important quantum interference phenomena have
been studied and measured in these mesoscopic systems in the presence 
of magnetic flux $\phi$~\cite{chand,mailly,keyser}. On the other hand, 
the future miniaturization of electronic devices have directed much 
more attention to characterize the structures, like an array of quantum 
dots, wires and rings at the sub-atomic level~\cite{lor,fuh,yan}. In the 
present paper we explore a theoretical study of the transport properties 
of a quantum ring placed between two macroscopic contacts in the presence 
of magnetic flux $\phi$. The ring may be treated as a chain of quantum 
dots or atoms. Electronic transport properties through a bridge system was 
first studied theoretically in $1974$~\cite{aviram}. The operation
of such two-terminal devices is due to an applied bias and the current 
passing across the junction is strongly nonlinear function of applied 
bias voltage. The complete knowledge of the conduction mechanism in this 
scale is not well established even today and its detailed description 
is quite complex. It has been verified that the transport properties 
in mesoscopic systems are strongly correlated with some quantum effects, 
like as quantization of energy levels, quantum interference of electron 
waves, etc. A quantitative understanding of the physical mechanisms 
underlying the operation of nanoscale devices remains a major challenge 
in nanoelectronics research.

Here, we reproduce an analytic approach based on the tight-binding model 
to investigate the electron transport properties in mesoscopic rings. 
There exist some {\em ab initio} methods for the calculation of 
conductance~\cite{yal,ven,xue,tay,der,dam}, yet the simple parametric 
approaches~\cite{muj1,muj2,sam,hjo,walc1,walc2} are needed for this 
calculation. The parametric study is much more flexible than that of 
the {\em ab initio} theories since the {\em ab initio} theories are 
computationally very expensive and here we concentrate only on the 
qualitative effects rather than the quantitative ones. This is why we 
restrict our calculations on the simple analytical formulation 
of the transport problem.

The organization of the present paper is as follows. In Section $2$, 
we present a brief description for the calculation of transmission 
probability and current through a finite size conducting system 
attached to two semi-infinite one-dimensional ($1$D) metallic electrodes 
by the use of Green's function method. Section $3$ focuses the results 
of conductance-energy ($g$-$E$) and current-voltage ($I$-$V$) 
characteristics for some typical isolated mesoscopic rings (both 
symmetrically and asymmetrically connected to the two electrodes), 
and at the end, we conclude our results in Section $4$. 

\section{A brief description of theoretical formulation}

\begin{figure}[ht]
{\centering \resizebox*{7.5cm}{1.5cm}{\includegraphics{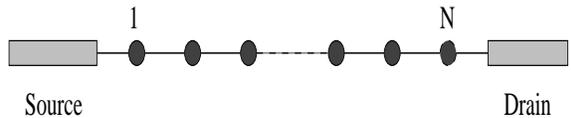}}\par}
\caption{Schematic representation of a one-dimensional conductor with 
$N$ atomic sites (filled circles) attached to two electrodes, viz, 
source and drain.}
\label{dot}
\end{figure}

We begin by referring to Fig.~\ref{dot}. A one-dimensional conductor 
with $N$ atomic sites (filled circles) connected to two semi-infinite 
$1$D metallic electrodes, viz, source and drain. The conducting system 
between the two electrodes
can be anything like, a mesoscopic ring, or an array of few quantum dots, 
or a single molecule, or an array of few molecules, etc. At the low 
voltage and low temperature limit, the conductance $g$ of the conductor 
can be written by using the Landauer conductance formula~\cite{tian,datta},
\begin{equation}
g=\frac{2e^2}{h}T
\label{land}
\end{equation}
where $T$ is the transmission probability of an electron through the 
conductor. This transmission probability can be expressed in terms of 
the Green's function of the conductor and the coupling of the conductor 
to the electrodes through the expression~\cite{tian,datta},
\begin{equation}
T={\mbox{Tr}} \left[\Gamma_S G_c^r \Gamma_D G_c^a\right]
\label{trans1}
\end{equation}
where $G_c^r$ and $G_c^a$ are the retarded and advanced Green's function 
of the conductor, respectively. $\Gamma_S$ and $\Gamma_D$ are the 
coupling terms of the conductor due to its coupling to the source and 
drain, respectively. For the complete system i.e., the conductor including
the two electrodes, the Green's function is defined as,
\begin{equation}
G=\left(\epsilon-H\right)^{-1}
\end{equation}
where $\epsilon=E+i\eta$. $E$ is the injecting energy of the source electron
and $\eta$ is a very small number which can be put to zero in the limiting
approximation. The above Green's function corresponds to the inversion of an
infinite matrix which consists of the finite conductor and two semi-infinite
electrodes. It can be partitioned into different sub-matrices that correspond
to the individual sub-systems.

The Green's function of the conductor can be effectively written 
as~\cite{tian,datta},
\begin{equation}
G_c=\left(\epsilon-H_c-\Sigma_S-\Sigma_D\right)^{-1}
\end{equation}
where $H_c$ is the Hamiltonian of the conductor. The single band 
tight-binding Hamiltonian for the conductor is written as,
\begin{equation}
H_c=\sum_i \epsilon_i c_i^{\dagger} c_i + \sum_{<ij>}t
\left(c_i^{\dagger}c_j + c_j^{\dagger}c_i \right)
\label{hamil1}
\end{equation}
where $\epsilon_i$'s are the on-site energies and $t$ is the 
nearest-neighbor hopping integral. Here $\Sigma_S=h_{Sc}^{\dagger} g_S h_{Sc}$
and $\Sigma_D=h_{Dc} g_D h_{Dc}^{\dagger}$ are the self-energy terms due to
the two electrodes. $g_S$ and $g_D$ correspond to the Green's functions for
the source and drain, respectively. $h_{Sc}$ and $h_{Dc}$ are the coupling
matrices and they will be non-zero only for the adjacent points in the 
conductor, $1$ and $N$ as shown in Fig.~\ref{dot}, and the electrodes 
respectively. The coupling terms $\Gamma_S$ and $\Gamma_D$ for the 
conductor can be calculated through the expression,
\begin{equation}
\Gamma_{\{S,D\}}=i\left[\Sigma_{\{S,D\}}^r-\Sigma_{\{S,D\}}^a\right]
\end{equation}
where $\Sigma_{\{S,D\}}^r$ and $\Sigma_{\{S,D\}}^a$ are the retarded and
advanced self-energies, respectively, and they are conjugate with each
other. Datta {\em et al.}~\cite{tian} have shown that the self-energies
can be expressed in the form,
\begin{equation}
\Sigma_{\{S,D\}}^r=\Lambda_{\{S,D\}}-i \Delta_{\{S,D\}}
\end{equation}
where $\Lambda_{\{S,D\}}$ are the real parts of the self-energies which
correspond to the shift of the energy eigenstates of the conductor and the
imaginary parts $\Delta_{\{S,D\}}$ of the self-energies represent the
broadening of these energy levels. Since this broadening is much larger than
the thermal broadening, we restrict our all calculations only at absolute 
zero temperature. By doing some simple algebra, the real and imaginary 
parts of the self-energies can be determined in terms of the coupling 
strength ($\tau_{\{S,D\}}$) between the conductor to the two electrodes, 
the injection energy ($E$) of the transmitting electron and the hopping 
strength ($v$) between nearest-neighbor sites in the electrodes. Thus the 
coupling terms $\Gamma_S$ and $\Gamma_D$ can be written in terms of the 
retarded self-energy as,
\begin{equation}
\Gamma_{\{S,D\}}=-2 {\mbox{Im}} \left[\Sigma_{\{S,D\}}^r\right]
\end{equation}
All the information regarding the conductor-to-electrode coupling are
included into the these two self energies and are analyzed through the use 
of Newns-Anderson chemisorption theory~\cite{muj1,muj2}. The detailed 
description of this theory is obtained in these two references.

Therefore, by calculating the self-energies, the coupling terms $\Gamma_S$ 
and $\Gamma_D$ can be easily obtained and then the transmission probability 
($T$) will be obtained from the expression as mentioned in Eq.~\ref{trans1}.

As the coupling matrices $h_{Sc}$ and $h_{Dc}$ are non-zero only for the 
adjacent points in the conductor, $1$ and $N$ as shown in Fig.~\ref{dot},
the transmission probability becomes,
\begin{equation}
T(E,V)=4\Delta_{11}^S(E,V) \Delta_{NN}^D(E,V)|G_{1N}(E,V)|^2
\label{trans2}
\end{equation}

The current passing through the conductor is depicted as a single-electron
scattering process between the two reservoirs of charge carriers and the
current-voltage relation is evaluated through the following
expression~\cite{datta},
\begin{equation}
I(V)=\frac{e}{\pi \hbar}\int \limits_{E_F-eV/2}^{E_F+eV/2} T(E,V) dE
\end{equation}
where $E_F$ is the equilibrium Fermi energy. For the sake of simplicity,
here we assume that the entire voltage is dropped across the
conductor-electrode interfaces and this assumption does not affect
significantly the current-voltage characteristics. Using the expression of
$T(E,V)$, given in Eq.~\ref{trans2}, the final form of $I(V)$ will be,
\begin{eqnarray}
I(V) &=& \frac{4e}{\pi \hbar}\int \limits_{E_F-eV/2}^{E_F+eV/2}
\Delta_{11}^S(E,V) \Delta_{NN}^D(E,V) \nonumber \\
& & \times |G_{1N}(E,V)|^2 dE
\label{curr}
\end{eqnarray}
Eqs.~\ref{land}, \ref{trans2} and \ref{curr} are the final working formule 
for the calculation of conductance $g$ and current-voltage ($I$-$V$) 
characteristics for any finite size conductor sandwiched between two 
electrodes.

Using the above formulation, we will describe the characteristic properties 
of electron transport for some mesoscopic rings. Throughout this article 
we fix the Fermi energy $E_F$ at $0$ and use the units $c=h=e=1$.

\section{Results and their interpretation}

Here we focus the conductance variation as a function of the energy and the
current-voltage characteristics of some mesoscopic rings, both symmetrically 
and asymmetrically connected to the two reservoirs. The results are 
investigated as functions of the effect of interference of electronic 
waves passing through different arms of the ring, ring-to-electrode 
coupling strength and magnetic flux. Due to the flux $\phi$, an 
additional phase difference appears between the electron waves 
transmitting through the two arms of the molecular ring, and accordingly, 
the tight-binding Hamiltonian Eq.~\ref{hamil1} gets modified by a phase 
factor. The single band tight-binding Hamiltonian that describes the 
mesoscopic ring in the presence of a magnetic flux can be written within 
the non-interacting picture in the form,
\begin{equation}
H_c=\sum_i \epsilon_i c_i^{\dagger} c_i + \sum_{<ij>}t
\left(e^{i\theta} c_i^{\dagger}c_j + e^{-i\theta} c_j^{\dagger}c_i \right)
\label{hamil2}
\end{equation}
where $\theta=2\pi \phi/N$, the phase factor due to the flux $\phi$ 
threaded by the ring with $N$ atomic sites and other symbols carry their 
usual meaning as in Eq.~\ref{hamil1}. Throughout the work we describe all 
the essential features of electron transport for the two limiting cases 
depending on the ring-to-electrode coupling strength. One is the so-called
weak coupling case, where the parameters are: $\tau_S=\tau_D=0.5$, $t=2.5$ 
and the other is the strong coupling case, where they are: 
$\tau_S=\tau_D=2$, $t=2.5$. The parameters $\tau_S$ and $\tau_D$ 
correspond to the coupling strengths of the ring to the source and drain, 
respectively.

\subsection{Ring sandwiched symmetrically between the two reservoirs}

\begin{figure}[ht]
{\centering \resizebox*{7.5cm}{3.65cm}{\includegraphics{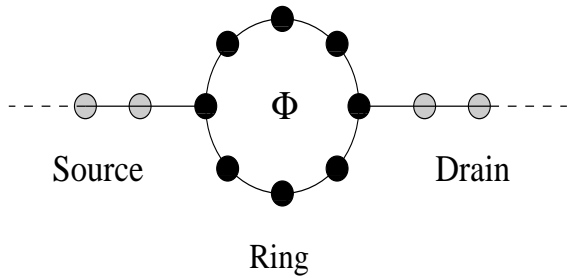}}\par}
\caption{Schematic representation of a mesoscopic ring, threaded by a 
magnetic flux $\phi$, sandwiched symmetrically between the two 
reservoirs.}
\label{ring1}
\end{figure}

\begin{figure}[ht]
{\centering \resizebox*{7.5cm}{5cm}
{\includegraphics{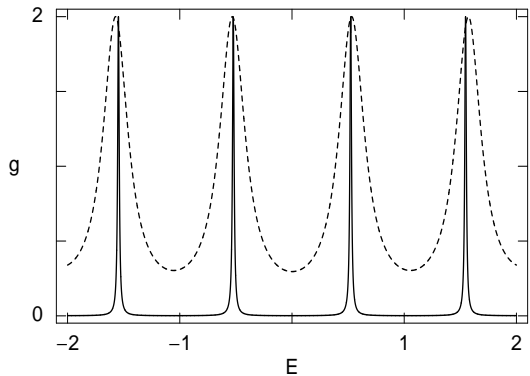}}\par}
\caption{Conductance $g$ as a function of the injecting electron energy $E$ 
for a symmetrically connected mesoscopic ring ($N=30$) in the absence of 
any magnetic flux, where the solid and dotted lines correspond to the 
results for the weak- and strong-coupling limits, respectively.}
\label{symcondphizero}
\end{figure}

The schematic view of a symmetrically connected mesoscopic ring is shown 
in Fig.~\ref{ring1}. Here the upper and lower arms contain equal number
of atomic sites. As illustrative example, in Fig.~\ref{symcondphizero}, 
we plot the conductance $g$ as a function of the injecting electron 
energy $E$ for a mesoscopic ring in the absence of any magnetic flux 
$\phi$, where the solid and dotted
lines denote the results for the weak- and strong-coupling cases, 
respectively. The size of the ring $N$ is fixed at $30$, where each arm 
of the ring contains $14$ number of atomic sites. Conductance shows 
oscillatory behavior with sharp resonant peaks for some particular energy 
values, while it almost vanishes for all other energies. At the
resonances, the conductance approaches the value $2$, and therefore, the
transmission probability $T$ goes to unity since we have the relation 
$g=2T$ from the Landauer formula with $e=h=1$. Such resonant peaks in the 
conductance spectrum coincide with the energy eigenvalues of the isolated 
mesoscopic ring and thus we can say that the conductance spectrum manifests 
itself the electronic structure of the ring. From this figure it is 
observed that the resonant peaks get substantial widths (dotted line), 
compared to the weak-coupling case, as long as the coupling strength of 
the ring to the electrodes is increased.

\begin{figure}[ht]
{\centering\resizebox*{7.5cm}{10cm}
{\includegraphics{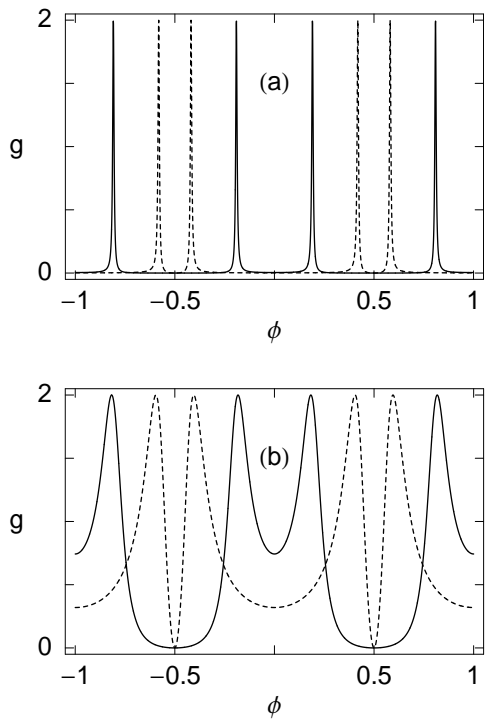}}\par}
\caption{$g$-$\phi$ spectra for a symmetrically connected mesoscopic ring 
($N=12$), where the solid and dotted curves correspond to the results
for the typical energies $E=0.5$ and $1.5$, respectively. (a) weak-coupling
limit and (b) strong-coupling limit.}
\label{symcondphinonzero}
\end{figure}

Now we discuss the effect of external magnetic flux, threaded by the 
ring, on electron transport.
In the presence of a magnetic flux, electron gains an additional phase, 
and accordingly, a constructive or destructive interference takes place 
after the electron propagation across the ring. Figure~\ref{symcondphinonzero} 
gives the variation of the conductance as a function of $\phi$ for a
particular energy, where (a) corresponds to the results for the 
weak-coupling limit and (b) denotes the results for the strong-coupling 
limit. Here we take the ring size $N=12$, as an illustrative example. 
The solid and dotted curves in the conductance spectra are associated
with the typical energies $E=0.5$ and $1.5$, respectively. It is observed 
that the conductance varies periodically with $\phi$ showing $\phi_0$ 
flux-quantum periodicity.

\begin{figure}[ht]
{\centering\resizebox*{7.5cm}{10cm}{\includegraphics{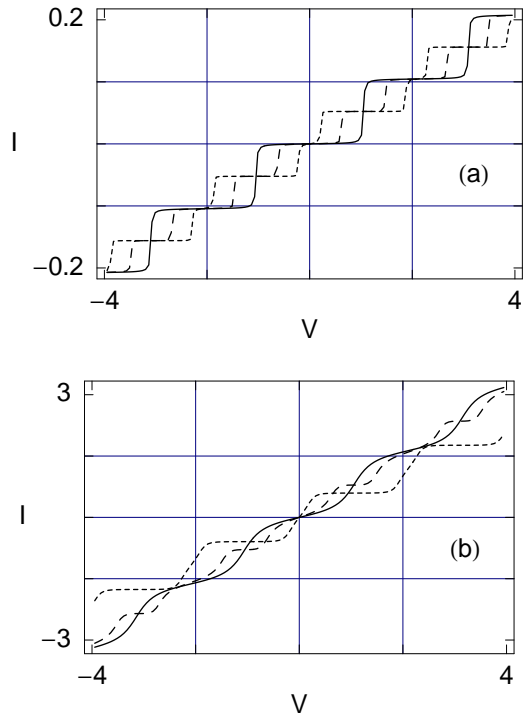}}\par}
\caption{$I$-$V$ spectra for a symmetrically connected mesoscopic ring 
($N=30$), where the solid, dotted and dashed curves correspond to the
results for $\phi=0$, $0.4$ and $0.8$, respectively. (a) weak-coupling
limit and (b) strong-coupling limit.}
\label{symcurr}
\end{figure}

The scenario of electron transfer through the ring becomes much more 
clearly visible by studying the current $I$ as a function of the applied 
bias voltage $V$. 
The variation of the current with the bias voltage is presented in 
Fig.~\ref{symcurr} for a mesoscopic ring with size $N=30$, where (a) and 
(b) correspond to the weak- and strong-coupling cases, respectively. 
The solid, dotted and dashed lines represent the results for $\phi=0$, 
$0.4$ and $0.8$, respectively. The current is evaluated by the 
integration procedure of the transmission function $T$, where the 
transmission function varies exactly similar to that of the conductance 
spectrum since the relation $g=2T$ exists from the Landauer conductance 
formula. The shape and height of the current steps depend on the width 
of the resonant peaks. In the weak-coupling limit, the current shows a 
staircase-like behavior with sharp steps, which is associated with 
discrete nature of the resonances. It is also noticed that in the 
presence of a magnetic flux the current shows more steps (dotted and 
dashed curves) compared to the current in 
the absence of any magnetic flux (solid line). This is due the fact that 
more resonant peaks appear in the conductance spectrum in the presence of
$\phi$. On the other hand, with the increase of the coupling strength, 
the current varies almost continuously with the applied bias voltage and 
achieves much larger values.

\subsection{Ring sandwiched asymmetrically between the two reservoirs}

\begin{figure}[ht]
{\centering \resizebox*{7.5cm}{3.65cm}{\includegraphics{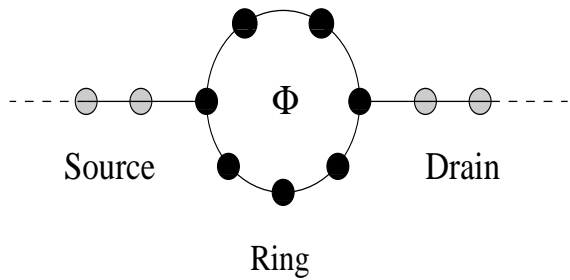}}\par}
\caption{Schematic view of a mesoscopic ring, threaded by a magnetic
flux $\phi$, attached asymmetrically between the two reservoirs.}
\label{ring2}
\end{figure}

The schematic view of an asymmetrically connected mesoscopic ring is 
shown in Fig.~\ref{ring2}, where the upper and lower arms contain 
unequal number of atomic sites. The idea of considering such a geometry 
is that, in this way the interference condition can be changed nicely 
and the effect of ring-electrode interface geometry can be clearly 
understood. It can be analyzed in this way. The electrons are carried 
from the source to drain
through the ring and the electron waves propagating along the two arms 
of the ring may suffer a phase shift among themselves. This is due to 
the result of quantum interference between the waves passing through the 
\begin{figure}[ht]
{\centering \resizebox*{7.5cm}{5cm}
{\includegraphics{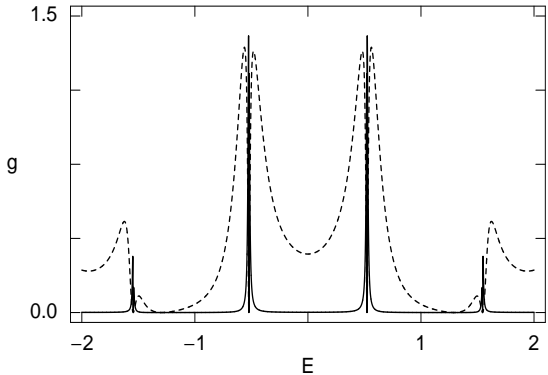}}\par}
\caption{Conductance as a function of the energy $E$ for an asymmetrically 
connected mesoscopic ring with $N=30$, where the upper and lower arms 
contain $8$ and $20$ atomic sites, respectively. $\phi$ is set at $0$. 
The solid and dotted curves represent the results for the weak- and 
strong-coupling limits, respectively.}
\label{asymcondphizero}
\end{figure}
two arms of the ring. As a result, the probability amplitude of the electron
across the ring may be strengthened or weakened, according to the standard
theory of quantum mechanics. In Fig.~\ref{asymcondphizero}, we plot the
conductance as a function of energy in the absence of any magnetic flux 
$\phi$ for an asymmetrically connected mesoscopic ring, where the upper and 
lower arms contain $8$ and $20$ atomic sites, respectively. The solid and 
dotted curves correspond to the results for the weak- and strong-coupling 
limits, respectively. It is observed that both for the strong- and 
weak-coupling cases the resonant peaks do not reach to $2$ anymore, i.e., 
transmission probability does not reach to unity. This is due to the 
effect of interference of the electronic waves 
\begin{figure}[ht]
{\centering \resizebox*{7.5cm}{10cm}
{\includegraphics{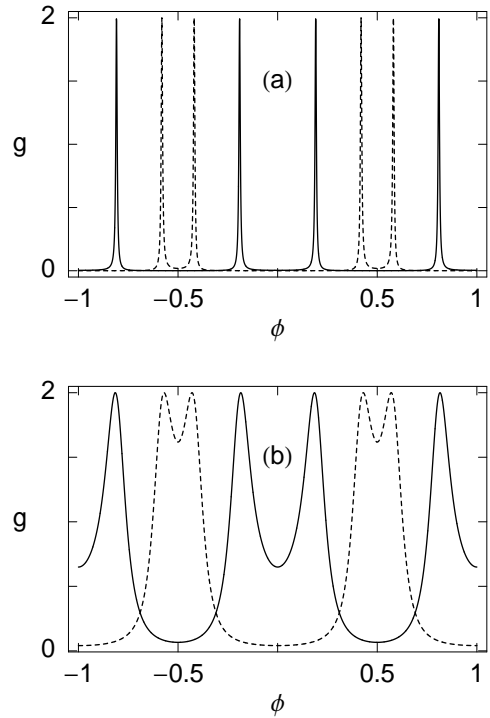}}\par}
\caption{$g$-$\phi$ characteristics for an asymmetrically connected
mesoscopic ring with $N=12$, where the upper and lower arms contain 
$2$ and $8$ atomic sites, respectively. The solid and dotted curves
represent the results for the typical energies $E=0.5$ and $1.5$, 
respectively. (a) weak-coupling limit and (b) strong-coupling limit.}
\label{asymcondphinonzero}
\end{figure}
propagating through the two arms of the ring. Like as in 
Fig.~\ref{symcondphizero}, here also the conductance peaks get substantial 
widths with the increase of coupling strength as shown by the dotted curve.

In the presence of $\phi$ in these asymmetrically connected rings, the 
conductance shows resonant and anti-resonant peaks, similar to that as 
shown in Fig.~\ref{symcondphinonzero}, but the widths and amplitudes of 
these peaks get modified significantly. Figure~\ref{asymcondphinonzero} 
shows the variation of the typical conductances as a function of flux 
$\phi$ for an asymmetrically connected ring, where the upper arm contains 
$2$ atomic sites and the lower arm contains $8$ atomic sites. The
typical conductances for the energy $E=0.5$ are shown by the solid
curves, while for the energy $E=1.5$ they are presented by the dotted
curves. The results for the weak- and strong-coupling cases are shown 
in (a) and (b), respectively. It is noticed that the typical conductance 
varies periodically with $\phi$ exhibiting $\phi_0$ flux-quantum 
periodicity.

\begin{figure}[ht]
{\centering\resizebox*{7.5cm}{9cm}{\includegraphics{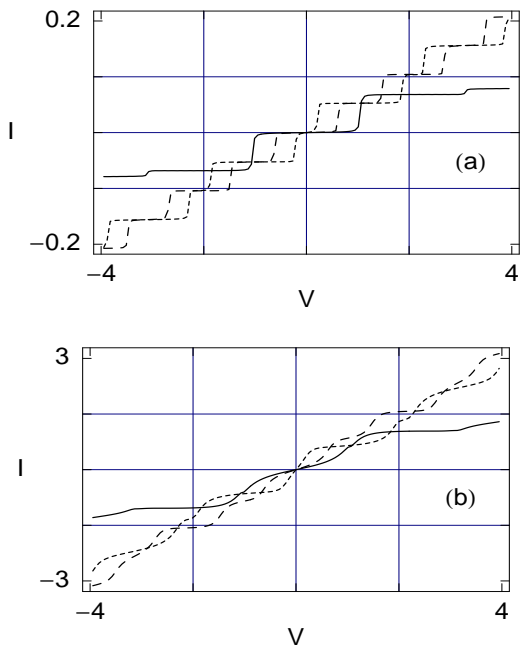}}\par}
\caption{$I$-$V$ spectra for an asymmetrically connected ring
with $N=30$, where the upper and lower arms have $8$ and $20$ atomic
site, respectively. The solid, dotted and dashed curves correspond to
the results for $\phi=0$, $0.4$ and $0.8$, respectively. (a) weak-coupling 
limit and (b) strong-coupling limit.}
\label{asymcurr}
\end{figure}

Now we discuss the effect of interference on the current-voltage 
characteristics for the asymmetrically connected ring. As representative 
example, in Fig.~\ref{asymcurr} we plot the current as a 
function of the bias voltage $V$ for an asymmetrically connected ring
with $N=30$, where the upper and lower arms contain $8$ and $20$ atomic
sites, respectively. The results for the weak- and strong-coupling limits
are shown in (a) and (b), respectively, where the solid, dotted and dashed 
curves correspond to the identical meaning as in Fig.~\ref{symcurr}. In 
the absence of any magnetic flux $\phi$, the current amplitudes
(solid curves) get reduced compared to the current amplitudes in the
presence of $\phi$ (dotted and dashed curves) both for the weak- and 
strong-coupling limits. This is due to the geometric effect of the 
asymmetrically connected ring. In the weak coupling case, the current 
shows staircase-like behavior with sharp steps, similar to the
symmetrically connected ring. Current shows more steps in the presence 
of $\phi$ than the current steps appear when $\phi$ becomes zero, since 
in the presence of $\phi$ more resonant peaks are obtained in the 
transmission spectrum. The current varies quite continuously for the 
strong-coupling limit and gets much higher values compared to the 
weak-coupling case, as expected from our previous discussion.

\section{Concluding remarks}

To summarize, we have addressed electron transport properties for some
typical isolated mesoscopic rings using the Green's function formalism.
We have introduced a parametric approach based on the tight-binding model 
to characterize the transport properties in such systems. This technique 
can be used to study the electronic transport in any complicated systems, 
like complicated organic molecules, which bridge the two reservoirs. Both 
the ring-electrode interface geometry and the magnetic flux threaded by
the ring have an important role on the electron transport since these 
two factors control the interference condition. We have also observed 
that the electron transport through the ring is significantly affected
by the ring-to-electrode coupling strength. All these features may provide
some basic inputs for fabrication of nanoscale devices.

\end{document}